\def\Journal#1#2#3#4{{#1} {\bf #2} (#4) #3}
 
\def\ARNPS{Annu. Rev. Nucl. Part. Sci.} 
\def\AandA{Astron. Astrophys.} 

\def\APJ{Astrophys. J.}

\def\CMP{Commn. Math. Phys.}

\def\CPC{Chin. Phys. C}
\def\CQG{Class. Quant. Grav.}
\def\EPJC{Eur. Phys. J. C}
\def\EPJP{Eur. Phys. J. Plus}

\def\IJMPA{Int. J. Mod. Phys. A}
\def\IJMPD{Int. J. Mod. Phys. D}

\def\JCAP{J. Cosmol. Astropart. Phys.}
\def\JHEP{J. High Energy Phys.}
\def\JETP{J. Exp. Theor. Phys}

\def\MPLA{Mod. Phys. Lett. A}
\def\MNRAS{Mon. Not. R. Astron. Soc.}

\def\NPB{Nucl. Phys. B}

\def\PLB{{Phys. Lett.} B}

\def\PPNP{Prog. Part. Nucl. Phys.}

\def\PRL{Phys. Rev. Lett.}
\def\PRD{Phys. Rev. D}

\def\PTP{Prog. Theor. Phys.}
\def\PTEP{Prog. Theor. Exp. Phys.}

\def\RPP{Rep. Prog. Phys.}

\def\SPP{SciPost Phys.}
\def\SCIENCE{Science}

\documentclass[3p,times,twocolumn]{elsarticle}

\usepackage{amssymb}
\usepackage{amsmath}
\usepackage{braket}
\usepackage{color}

\biboptions{sort&compress}

\usepackage[figuresright]{rotating}

\begin{document}

\begin{frontmatter}



\title{Constraining scotogenic dark matter and primordial black holes using \\  induced gravitational waves}



\author{Teruyuki Kitabayashi}
\ead{teruyuki@tokai-u.jp}
\address{Department of Physics, Tokai University, 4-1-1 Kitakaname, Hiratsuka, Kanagawa, 259-1292, Japan}

\begin{abstract}
The lightest $Z_2$ odd particle in the scotogenic model, referred to as scotogenic dark matter (DM), is a widely studied candidate for DM. This scotogenic DM is generated through well-known thermal processes as well as via the evaporation of primordial black holes (PBHs). Recent reports suggested that the curvature fluctuations of PBHs during an epoch dominated by these entities in the early universe can serve as the source of so-called induced gravitational waves (GWs). In this study, we demonstrate that stringent constraints on the mass of scotogenic DM and PBHs can be obtained through the detection of induced GWs using future detectors.
\end{abstract}

\begin{keyword}
Scotogenic dark matter \sep Primordial black holes \sep Induced gravitational waves


\end{keyword}
\end{frontmatter}


\section{Introduction\label{section:introduction}}
The existence of dark matter (DM) and the minute scale of neutrino masses have been a longstanding mystery in cosmology and particle physics. The scotogenic model offers a unified explanation for the presence of DM and the origin of tiny neutrino masses \cite{Ma2006PRD}. In this model, neutrino masses are generated from one-loop interactions mediated by a DM candidate. The DM in the scotogenic model (scotogenic DM) is generated due to the thermal processes in the early universe, which has garnered extensive attention in the literature. \cite{Suematsu2009PRD,Suematsu2010PRD,Kubo2006PLB,Hambye2007PRD,Farzan2009PRD,Farzan2010MPLA,Farzan2011IJMPA,Kanemura2011PRD,Schmidt2012PRD,Faezan2012PRD,Aoki2012PRD,Hehn2012PLB,Bhupal2012PRD,Bhupal2013PRD,Law2013JHEP,Kanemura2013PLB,Hirsch2013JHEP,Restrepo2013JHEP,Ho2013PRD,Lindner2014PRD,Okada2014PRD89,Okada2014PRD90,Brdar2014PLB,Toma2014JHEP,Ho2014PRD,Faisel2014PRD,Vicente2015JHEP,Borah2015PRD,Wang2015PRD,Fraser2016PRD,Adhikari2016PLB,Ma2016PLB,Arhrib2016JCAP,Okada2016PRD,Ahriche2016PLB,Lu2016JCAP,Cai2016JHEP,Ibarra2016PRD,Lindner2016PRD,Das2017PRD,Singirala2017CPC,Kitabayashi2017IJMPA,AbadaJHEP2018,Baumholzer2018JHEP,Ahriche2018PRD,Hugle2018PRD,Kitabayashi2018PRD,Reig2019PLB,Boer2020PRD,Ahriche2020PRD,Faisel2014PLB,Cacciapaglia2021JHEP}.

The evaporation of primordial black holes (PBHs) serve as a potential source for the production of scotogenic DM \cite{Kitabayashi2021IJMPA,Kitabayashi2022PTEP1}. These PBHs are among the types of black holes formed in the early universe \cite{Carr1975APJ,Carr2020ARNPS,Carr2021RPP,Auffinger2023PPNP}. PBHs emit particles via the Hawking radiation \cite{Hawking1975CMP}. Owing to the Hawking radiation induced by gravity, PBHs are expected to produce some additional particles in addition to standard model particles, independent of their nongravitational interactions \cite{Bell1999PRD,Green1999PRD,Khlopov2006CQG,Baumann2007arXiv,Dai2009JCAP,Fujita2014PRD,Allahverdi2018PRD,Lennon2018JCAP,Morrison2019JCAP,Hooper2019JHEP,Masina2020EPJP,Baldes2020JCAP,Bernal2021JCAP,Gondolo2020PRD,Bernal2021PLB,Datta2021JCAP,Chaudhuri2021JETP,Sandick2021PRD,Cheek2022PRD,Baker2022SciPostPhys,Das2021arXiv,Auffinger2021EPJP,Kitabayashi2022PTEP2,Kitabayashi2022IJMPA}. Thus, given the coexistence of the scotogenic DM and PBHs in the universe, the scotogenic DM should be produced by thermal processes as well as PBH evaporation. 

Moreover, PBHs have been identified as a potential source of DMs and gravitational waves (GWs) \cite{Tomita1967PTP,Matarrese1993PRD,Matarrese1994PRL,Matarrese1998PRD,Noh2004PRD,Carbone2005PRD,Nakamura2007PTP,Ananda2007PRD,Baumann2007PRD,Osano2007JCAP,Espinosa2018JCAP,Kohr2018PRD,Domenech2020IJMPD,Domenech2020JCAP,Wang2023arXiv,Choudhury2014PLB,Choudhury2023arXiv}. In the last stage of the inflation, immediately after horizon reentry, the primordial fluctuations in the curvature perturbation can act as a source of induced GWs. Furthermore, the PBH formation is invariably accompanied by GWs. The induced GWs have garnered considerable research interest as they could be detected by the currently developing future GW observations such as Hanford-Livingstone-Virgo-KAGRA (HLVK, or aLIGO-aVirgo-KAGRA) \cite{LIGO2010CQG,LIGO2015CQG,VIRGO2015CQG,KAGRA2012CQG}, Einstein Telescope (ET) \cite{Punturo2010CQG,Maggiore2020JCAP}, Cosmic Explorer (CE) \cite{LIGO2017CQG}, Laser Interferometer Space Antenna (LISA) \cite{LISA2017arXiv}, Deci-Hertz Interferometer Gravitational-Wave Observatory (DECIGO) \cite{Seto2001PRL,Kawamura2006CQG} and Big Bang Observer (BBO) \cite{Crowder2005PRD,Harry2006CQG}.

Recent reports indicated that GWs could be induced by sources other than the primordial fluctuation \cite{Inomata2019JCAP,Inomata2020PRD,Papanikolaou2021JCAP}. Immediately after formation, PBHs are randomly distributed in space according to approximate Poisson statistics. Although the PBH gas behaves as pressureless dust, the spatially inhomogeneous distribution leads to isocurvature density fluctuations. At a later time, if the initial fraction of PBHs is sufficiently large, they would eventually overcome the energy density of the radiation component, thereby dominating the universe. During the transition from an early radiation-dominated epoch to a PBH-dominated (in other words, an early matter-dominated) epoch, the initial isocurvature fluctuations are transmuted into curvature perturbations, which consequently act as sources for secondary GWs \cite{Inomata2019JCAP,Inomata2020PRD,Inomata2019PRD,Papanikolaou2021JCAP,Domenech2021JCAP,Domenech2021PLB,Borah2023JHEP,KawasakiArXiv2023}.

In a study reported in Ref. \cite{Borah2023JHEP}, constraints on heavy particles serving as DM in the seesaw mechanism and PBHs are inferred using induced GWs in the PBH-dominated epoch. The seesaw mechanism is a widely acknowledged model for the generation of diminutive neutrino masses \cite{Minkowski1977PLB,Yanagida1979KEK,Gell-Mann1979,Glashow1979,Mohapatra1980PRL}. Currently, the seesaw mechanism and the scotogenic model represent the most extensively researched paradigms for neutrino mass production.

In this study, we postulate that scotogenic DM constitutes the DM and that a PBH-dominated epoch was existent in the early universe. By employing a methodology analogous to that reported in Ref.\cite{Borah2023JHEP}, we estimate the constraints on the masses of scotogenic DM and PBHs based on induced GWs \footnote{The relation between the scotogenic model and GWs from first order phase transitions are studied in Refs. \cite{Borah2020JCAP,Bonilla2023arXiv}. On the contrary, in this paper, we show the relation between the scotogenic model, PBHs and the induced GWs.}. Although Ref. \cite{Borah2023JHEP} suggests DM production only through PBH evaporation, this study additionally incorporates the freeze-out mechanism for scotogenic DM production. We demonstrate that future GW detectors could impose rigorous constraints on the masses of scotogenic DM and PBHs.

The remainder of this article is organized as follows. In Sec. \ref{section:scotogenic_and_PBH}, we offer a short review of scotogenic DM and PBHs, emphasizing DM production via the freeze-out mechanism and PBH evaporation, which is one of the primary topics. In addition, we focus on the necessary condition of the initial density of PBHs to  establish a PBH-dominated epoch and maintain consistency with Big Bang Nucleosynthesis (BBN). In Sec. \ref{section:GWs}, we present the stringent constraints on the mass of scotogenic DM and PBHs deduced from induced GWs. The article concludes with a summary of the key findings in Sec. \ref{section:summary}.

\section{Scotogenic dark matter and primordial black holes \label{section:scotogenic_and_PBH}}
\subsection{Scotogenic model}
In the scotogenic model, three new Majorana $SU(2)_L$ singlets $N_k$ $(k=1,2,3)$ with mass $M_k$ and one new scalar $SU(2)_L$ doublet $(\eta^+,\eta^0)$ are introduced into the standard model \cite{Ma2006PRD}. The Lagrangian and scalar potential relevant for this study are expressed as
\begin{eqnarray}
\mathcal{L} &=& Y_{\alpha k} (\bar{\nu}_{\alpha L} \eta^0 - \bar{\ell}_{\alpha L} \eta^+) N_k + \frac{1}{2}M_k \bar{N}_k N^C_k + h.c.,
\nonumber \\
V &=& \frac{1}{2}\lambda (\phi^\dagger \eta)^2 + h.c.,
\label{Eq:L_V}
\end{eqnarray}
where $L_\alpha=(\nu_\alpha, \ell_\alpha)$ $(\alpha=e,\mu,\tau)$ denotes the left-handed lepton doublet and $\phi=(\phi^+, \phi^0)$ indicates the standard Higgs doublet. The new particles are odd under exact $Z_2$ symmetry, and although the tree level neutrino mass should disappear, they acquire masses via one-loop interactions. 

The flavor neutrino mass matrix $M$ is obtained as 
\begin{equation}
M_{\alpha\beta} = \sum_{k=1}^3  \frac{\lambda v^2 Y_{\alpha k}Y_{\beta k}M_k}{16\pi^2(m^2_0-M^2_k)}\left(1-\frac{M^2_k}{m^2_0-M^2_k}\ln\frac{m_0^2}{M^2_k} \right),
\end{equation}
where $m_0^2 = \frac{1}{2}(m_R^2+m_I^2)$ and $v$, $m_R$, $m_I$ denote vacuum expectation value of the Higgs field, the masses of $\sqrt{2} {\rm Re}[\eta^0]$ and $\sqrt{2} {\rm Im}[\eta^0]$, respectively. 

Given that the lightest $Z_2$ odd particle is stable, it is a suitable candidate for DM. As reported earlier, if the lightest singlet fermion $N_1$ is the DM and $N_1$ is nearly degenerated with the subsequently lighter singlet fermions, the observed relic abundance of DM and the upper limit of the branching ratio of the $\mu \rightarrow e \gamma$ process can be simultaneously consistent with the predictions from the scotogenic model \cite{Suematsu2009PRD,Suematsu2010PRD}. Thus, we assume that the lightest singlet fermion $N_1$ is the DM and $M_1 \simeq M_2  < M_3<m_0$. Hereafter, we used a notation of $m_{\rm DM} = M_1$.

The branching ratio of flavor violating $\mu \rightarrow e \gamma$ processes is expressed as \cite{Kubo2006PLB}
\begin{equation}
{\rm BR}(\mu \rightarrow e \gamma)=\frac{3\alpha_{\rm em}}{64\pi(G_{\rm F} m_0^2)^2}\left| \sum_{k=1}^3 Y_{\mu k}Y_{e k}^* F \left( \frac{M_k}{m_0}\right) \right|^2,
\end{equation}
where $\alpha_{\rm em}$ denotes the fine-structure constant, and $G_{\rm F}$ denotes the Fermi coupling constant and
\begin{eqnarray}
F(x)=\frac{1-6x^2+3x^4+2x^6-6x^4 \ln x^2}{6(1-x^2)^4}.
\end{eqnarray}

\subsection{DM production by freeze-out mechanism \label{sec:FO}}
The scotogenic DM, lightest $Z_2$ odd particle $N_1$, can reach thermal equilibrium with the thermal bath particles in the early universe through annihilation and pair production processes. When the temperature in the thermal bath particles drops below the mass of scotogenic DM, the density of scotogenic DM is exponentially suppressed by Boltzmann factor until $\Gamma \lesssim H$, where $\Gamma$ and $H$ denote the reaction rate of annihilation and the Hubble parameter, respectively. Accordingly, the comoving number density of scotogenic DM remains constant, which is the so-called freeze-out mechanism. The relic abundance of scotogenic DM produced by the freeze-out mechanism is estimated to be \cite{Griest1991PRD,KolbTurner1991}:
\begin{eqnarray}
\Omega_{\rm FO} h^2 = \frac{1.07\times 10^9 x_{\rm FO}}{g_\ast^{1/2} M_{\rm Pl} (a_{\rm eff}+3b_{\rm eff}/x_{\rm FO})},
\end{eqnarray}
where $h=0.674$ denotes the dimensionless Hubble constant \cite{Planck2020AA}, $g_\ast$ indicates the effective degrees of freedom of the relativistic particles for energy density, $M_{\rm Pl} = 1.22 \times 10^{19} \ {\rm GeV} = 2.18\times 10^{-5}\  {\rm g}$ represents the Plank Mass, $x_{\rm FO} = m_{\rm DM}/T_{\rm FO} \simeq 25$ denotes the freeze-out temperature, and \cite{Suematsu2009PRD,Suematsu2010PRD,Kubo2006PLB}
\begin{eqnarray}
a_{\rm eff} &=& \frac{a_{11}}{4}+\frac{a_{12}}{2}+\frac{a_{22}}{4},  \nonumber \\
b_{\rm eff} &=& \frac{b_{11}}{4}+\frac{b_{12}}{2}+\frac{b_{22}}{4},
\label{Eq:aeff_beff}
\end{eqnarray}
with
\begin{eqnarray}
a_{ij}&=& \frac{1}{8\pi}\frac{m_{\rm DM}^2}{(m_{\rm DM}^2+m_0^2)^2} \sum_{\alpha\beta}(Y_{\alpha i} Y_{\beta j} - Y_{\alpha j} Y_{\beta i})^2, \nonumber \\
b_{ij}&=&\frac{m_0^4-3m_0^2m_{\rm DM}^2-m_{\rm DM}^4}{3(m_{\rm DM}^2+m_0^2)^2}a_{ij}  \nonumber \\
&& +  \frac{1}{12\pi}\frac{m_{\rm DM}^2(m_{\rm DM}^4+m_0^4)}{(m_{\rm DM}^2+m_0^2)^4}  \sum_{\alpha\beta}Y_{\alpha i} Y_{\alpha j} Y_{\beta i} Y_{\beta j}. \nonumber \\
\label{Eq:a_b}
\end{eqnarray}
%

\subsection{Initial density of PBHs}
We assume that PBHs are produced by large density perturbations generated from an inflation, and the PBHs form during the radiation-dominated epoch with a monochromatic mass function \cite{Garcia-Bellido1996PRD,Kawasaki1998PRD,Yokoyama1998PRD,Kawasaki2006PRD,Kawaguchi2008MNRAS,Kohri2008JCAP,Drees2011JCAP,Lin2013PLB,Linde2013PRD}. The dimensionless parameter can be expressed as
\begin{eqnarray}
\beta = \frac{\rho_{\rm PBH}(T_{\rm in})}{\rho_{\rm rad}(T_{\rm in})},
\label{Eq:beta}
\end{eqnarray}
This is introduced to represent the initial energy density of PBHs at the instant of its formation, $\rho_{\rm PBH}(T_{\rm in})$, where $\rho_{\rm rad}(T_{\rm in})$ indicates the energy density of radiation and $T_{\rm in}$ denotes the temperature of the universe at PBH formation time.

For the PBH evaporation to occur in the PBH-dominated epoch, the initial density of PBHs $\beta$ should be greater than the critical density \cite{Fujita2014PRD,Hamdan2018MPLA,Masina2020EPJP,Baldes2020JCAP}, expressed as follows:
\begin{eqnarray}
\beta_{\rm c} = \frac{T_{\rm evap}}{T_{\rm in}} \simeq  0.129 \left(\frac{g_*(T_{\rm BH})}{106.75}\right)^{1/2} \left(\frac{M_{\rm PBH}} {M_{\rm Pl}}\right)^{-1},
\label{Eq:beta_c}
\end{eqnarray}
where $T_{\rm evap}$ and $M_{\rm PBH}$ denote the temperature of the universe at PBH evaporation and the initial PBH mass, respectively.

Furthermore, as any additional relativistic particle species during radiation-dominated epoch at BBN should contribute to at most 20\% of the total energy density, the upper limit of the initial density of PBHs can be derived as \cite{Planck2020AA,Caplini2018CQG,Domenech2021JCAP}
\begin{eqnarray}
\beta \lesssim 1.1 \times 10^{-6} \left( \frac{M_{\rm PBH}}{10^4 {\rm g}} \right)^{-17/24}.
\label{Eq:betaUpperBBN}
\end{eqnarray}
%

\subsection{DM production by PBH evaporation \label{sec:FO}}
Scotogenic DMs are generated from the Hawking radiation of PBHs when the mass of scotogenic DM is lower than the Hawking temperature of the PBH. The density of scotogenic DM produced by the Hawking radiation of PBHs in the PBH-dominated epoch can be derived as \cite{Kitabayashi2021IJMPA,Kitabayashi2022PTEP1,Fujita2014PRD,Hamdan2018MPLA,Masina2020EPJP,Baldes2020JCAP}.
\begin{eqnarray}
\Omega_{\rm PBH}  h^2 &\simeq&  1.09 \times 10^7  \left(\frac{g_*(T_{\rm BH})}{106.75} \right)^{1/4}  \nonumber \\
&&\times \frac{3}{4} \frac{g_{\rm DM}  }{g_*(T_{\rm BH})}  \left( \frac{m_{\rm DM}}{{\rm GeV}} \right) \left( \frac{M_{\rm Pl}}{M_{\rm PBH}} \right)^{1/2}.
\label{Eq:OmegaPBHh2_MD_T>m} 
\end{eqnarray}

If the PBH evaporates after the freeze-out of the scotogenic DM, the scotogenic DM produced from the PBHs may contribute to the final relic abundance of the DM \cite{Gondolo2020PRD}. In the PBH-dominated epoch, the entropy production via the evaporation of PBHs causes a dilution of the freeze-out-origin scotogenic DM \cite{Kitabayashi2021IJMPA,Kitabayashi2022PTEP1,Fujita2014PRD,Hamdan2018MPLA,Masina2020EPJP,Baldes2020JCAP,KolbTurner1991}. The final relic abundance of the scotogenic DM is derived as
\begin{eqnarray}
\Omega_{\rm DM} h^2 = \alpha^{-1} \Omega_{\rm FO}h^2 + \Omega_{\rm PBH}h^2.
\label{Eq:OmegaDM_OmegaFO_OmegaPBH}
\end{eqnarray}
The factor $\alpha$ denotes the ratio of the entropy prior and after PBH evaporation, expressed as
\begin{eqnarray}
\alpha = \beta \frac{g_\ast(T_{\rm in})}{g_\ast(T_{\rm evap})}\frac{g_{\ast s}(T_{\rm evap})}{g_{\ast s}(T_{\rm in})}\frac{T_{\rm in}}{T_{\rm evap}} \simeq \frac{\beta}{\beta_{\rm c}},
\end{eqnarray}
where we use the relation $g_\ast(T) \simeq g_{\ast s}(T)$ for high temperature, with $g_{\ast s}$ denoting the effective degrees of freedom of the relativistic particles for entropy density.

\section{Gravitational waves \label{section:GWs}}

\subsection{Induced GWs in PBH-dominated epoch}
As discussed in the introduction, although the PBH gas in the early universe on average behaves like pressure-less dust, its spatially inhomogeneous distribution leads to isocurvature density fluctuations. Should the initial fraction of PBHs be sufficiently large, a transition from the early radiation-dominated epoch to the PBH-dominated epoch would occur, converting the early isocurvature component into curvature perturbations. These curvature perturbations serve as sources of secondary GWs. In case of large density fluctuations on small scales, large GWs are induced \cite{Inomata2019PRD,Papanikolaou2021JCAP,Domenech2021JCAP,Domenech2021PLB,Borah2023JHEP}.

The present-day amplitude (energy density) of the induced GWs is expressed as \cite{Domenech2021JCAP,Borah2023JHEP}.
\begin{eqnarray}
\Omega_{\rm GW} \simeq \Omega_{\rm GW}^{\rm peak} \left( \frac{f}{f^{\rm peak}} \right)^{11/3} \Theta \left( f^{\rm peak}-f \right),
\label{Eq:OmegaGW}
\end{eqnarray}
where 
\begin{eqnarray}
\Omega_{\rm GW}^{\rm peak} \simeq 2\times 10^{-6} \left( \frac{\beta}{10^{-8}} \right)^{16/3} \left( \frac{M_{\rm PBH}}{10^7 {\rm g}} \right)^{34/9},
\label{Eq:OmegaGWpeak}
\end{eqnarray}
denotes the peak amplitude and 
\begin{eqnarray}
f^{\rm peak} \simeq 1.7\times 10^3  \left( \frac{M_{\rm PBH}}{10^4 {\rm g}} \right)^{-5/6} \ {\rm Hz},
\label{Eq:fpeak}
\end{eqnarray}
denotes the peak frequency. The Heaviside function $\Theta$ is due to the ultraviolet cutoff in the Poisson spectrum of the density fluctuations of PBH \cite{Domenech2021JCAP,Borah2023JHEP}.

We comment that the amplitude of the induced GWs is extremely sensitive to the PBH mass spectrum \cite{Inomata2020PRD,Borah2023JHEP,Domenech2021JCAP,Papanikolaou2022JCAP}. The findings in this study are contingent upon a monochromatic mass spectrum, and might be subject to modification if an extended mass function is considered.

\subsection{Constraints without GWs}
\begin{figure}[t]
\begin{center}
\includegraphics[scale=1.0]{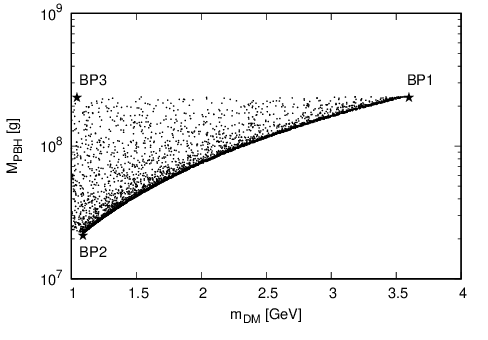}
\includegraphics[scale=1.0]{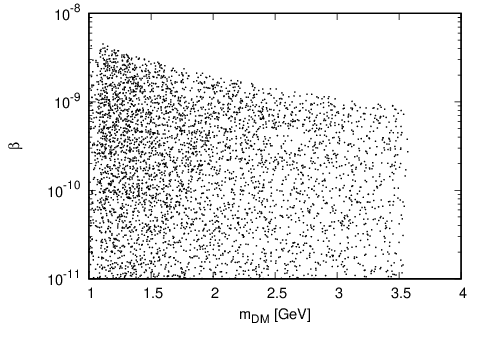}
\caption{Top (bottom) panel exhibits the allowed mass of scotogenic DM, $m_{\rm DM}$, and the allowed mass of PBH, $M_{\rm PBH}$, (allowed initial PBH density $\beta$) from particle physics experiments and CMB observations. BP1, BP2, and BP3 are benchmark points that will be used later.}
\label{Fig:mdm_mpbh_beta} 
\end{center}
\end{figure}
\begin{figure*}[t]
\begin{center}
\includegraphics[scale=0.9]{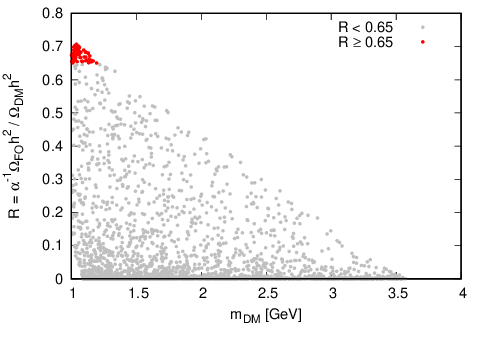} 
\includegraphics[scale=0.9]{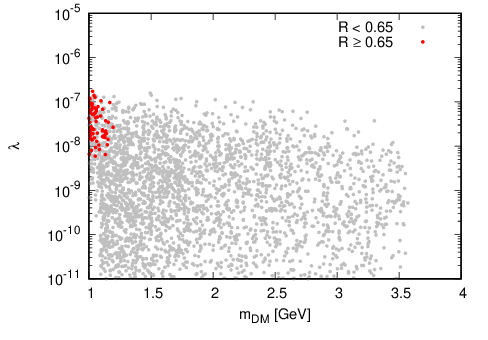} \\
\includegraphics[scale=0.9]{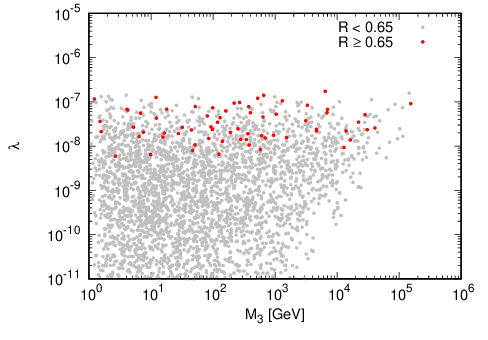}
\includegraphics[scale=0.9]{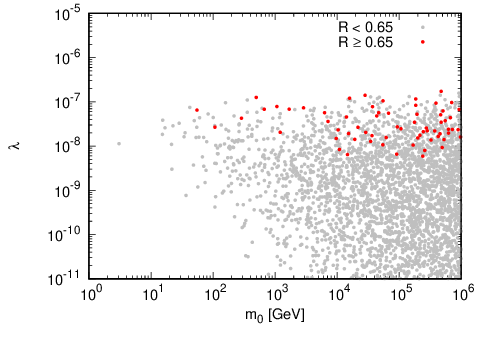} 
\caption{\color{black}Dependence the ratio $R = \alpha^{-1}\Omega_{\rm FO}h^2 / (\Omega_{\rm DM}h^2)$ on the scotogenic model parameters $m_{\rm DM}$,  $\lambda$, $M_3$ and $m_0$. The gray and red dots indicate the points which are satisfied with $R < 0.65$, and $R \ge 0.65$, respectively. }
\label{Fig:fig_HighOmegaScoto} 
\end{center}
\end{figure*}

First, we estimate the constraint on the mass of scotogenic DM, $m_{\rm DM}$, and the initial PBH mass, $M_{\rm PBH}$, without considering GWs through numerical calculations

In the numerical calculations, we use the best-fit values of neutrino parameters in Ref. \cite{Esteban2020JHEP}. For simplicity, the normal mass ordering for the neutrinos $m_1 < m_2 < m_3$ is assumed and the mass of the lightest neutrino is taken as
\begin{eqnarray}
m_1= 0.001 - 0.1 \ {\rm eV}.
\end{eqnarray}
As the relic density of the scotogenic DM depends weakly on CP-violating phases, the Majorana CP phases are neglected \cite{Boer2020PRD}. For the remaining four parameters in the scotogenic model, $\{m_{\rm DM}, M_3, m_0, \lambda\}$.
\color{black}
In this study, the lightest $Z_2$ odd fermion is DM. In this case, the annihilation cross-section of scotogenic DM is suppressed due to the its Majorana nature. Thus, the indirect detection experiments are hard to have positive signals \cite{Kubo2006PLB}. On the other hand, the DM scattering with nuclei via the one-loop exchange of a photon, $Z^0$ and Higgs boson may also be naturally suppressed by the loop factor; however, under some specific situation, we can expect the signal from current and upcoming experiments \cite{Schmidt2012PRD,Ibarra2016PRD}. The lower bound on the DM mass is taken to be 1 GeV. These experimental bound does not allow take smaller mass of DM while allowing a wide range of PBH masses. For parameters in the scotogenic model, we set the commonly used following range \cite{Kubo2006PLB,Ibarra2016PRD,Lindner2016PRD}:  
\color{black}
\begin{eqnarray}
&&1.0  \ {\rm GeV} \  \le m_{\rm DM}, M_3, m_0 \le 1.0 \times10^6 \ {\rm GeV}, \nonumber \\
&&1.0 \times 10^{-11} \le \lambda \le 1.0 \times 10^{-6}.
\label{Eq:parameters}
\end{eqnarray}
%
\color{black}
In this model, the Yukawa coupling $Y$ can be obtained using the Casas-Ibarra parametrization \cite{Casas2001NPB}.  

For PBH evaporation after the freeze-out of the DM, the lower limit on the initial PBH is obtained \cite{Gondolo2020PRD,Kitabayashi2021IJMPA}. In addition, the upper limit on the initial PBH mass is derived based on BBN \cite{Fujita2014PRD,Kohri1999PRD,Kawasaki2000PRD}. The upper and lower limits of the initial PBH mass are conservatively set as follows:
\begin{eqnarray}
2 \times 10^{12} \left( \frac{{\rm GeV}}{m_{\rm DM}} \right)^{2/3} \le \frac{M_{\rm PBH}}{M_{\rm Pl}} \le 2 \times 10^{13}.
\end{eqnarray}

The predicted physical quantities should be satisfied with the following conditions obtained from particle physics experiments and the cosmic microwave background (CMB) observations, 
\begin{itemize}
\item $|M_{ee}| < 0.066 \ {\rm eV}$ from neutrinoless double beta decay experiment \cite{Capozzi2020PRD,GERDA2019Science}.
\item ${\rm BR}(\mu\rightarrow e\gamma) \le 4.2\times 10^{-13}$ from the MEG experiment  \cite{MEG2016EPJC}.
\item $\sum m_i < 0.12 \ {\rm eV}$ and $\Omega_{\rm DM} h^2 = 0.12$ from CMB observations \cite{Planck2020AA}.
\end{itemize}

The top (bottom) panel of Fig. \ref{Fig:mdm_mpbh_beta} displays the allowed mass of scotogenic DM $m_{\rm DM}$ and the allowed mass of PBH $M_{\rm PBH}$ (allowed initial PBH density $\beta$) from particle physics experiments and CMB observations. BP1, BP2, and BP3 indicate the benchmark points discussed later. The constraint on the mass of scotogenic DM and the initial PBH mass are obtained as
\begin{eqnarray}
\label{Eq:constraint_wo_GWs}
&&
\begin{aligned}
1.00 \lesssim  m_{\rm DM} [{\rm GeV}] \lesssim 3.57, \\
2.22  \lesssim  M_{\rm PBH}  [10^{7} {\rm g}] \lesssim 23.5,  
\end{aligned} \\
&& \qquad {\rm (Particle \ physics \ experiments \ and \ CMB)}. \nonumber
\end{eqnarray}
The allowed $m_{\rm DM}$ is restricted to an extremely narrow range in the order of GeV. Notably, the allowed range of $M_{\rm PBH}$ is not substantially wide. The upper bound displayed in the bottom panel in Fig. \ref{Fig:mdm_mpbh_beta} emerges from the BBN constraints expressed in Eq.(\ref{Eq:betaUpperBBN}). The constraints on the masses expressed in Eq. (\ref{Eq:constraint_wo_GWs}) is consistent with the results in Ref. \cite{Kitabayashi2021IJMPA,Kitabayashi2022PTEP1}.

\color{black}
As shown in Eq. (\ref{Eq:OmegaDM_OmegaFO_OmegaPBH}), there are two contributions to the DM abundance. One is the freeze-out origin particle, $\alpha^{-1}\Omega_{\rm FO}h^2$, and the other is the particle coming from PBH evaporation, $\Omega_{\rm PBH}h^2$. One may wonder that which one is dominant in the total DM abundance. In general, the contribution to the relic abundance of DM from the PBH evaporation does not depend on a specific DM model. Therefore, if PBH contribution dominates in the total DM energy density, it seems difficult to distinguish the scotogenic model from other numerous DM models.

To clarify this issue, the ratio
\begin{eqnarray}
R = \frac{\alpha^{-1}\Omega_{\rm FO}h^2}{\Omega_{\rm DM}h^2} 
\end{eqnarray}
is useful. The undesired situation may be avoided if $R \geq 0.65$. In this case, over 65\% of total DM abundance is made of freeze-out origin DM. (Although 0.65 is an ad-hoc criteria, we use this criteria for our purpose).

The main parameters in the scotogenic model in this study are $m_{\rm DM}$,  $\lambda$, $M_3$ and $m_0$ as shown in Eq. (\ref{Eq:parameters}). We show the dependence the ratio $R$ on these parameters in Fig. \ref{Fig:fig_HighOmegaScoto}. The upper-left panel of Fig. \ref{Fig:fig_HighOmegaScoto} depicts $R$ vs. $m_{\rm DM}$. The gray and red dots indicate the points which are satisfied with $R < 0.65$, and $R \ge 0.65$, respectively.  Since we consider the case $R \ge 0.65$ to be favorable, the relatively light DM (red dot) is desired to distinguish the scotogenic model from other numerous DM models. The upper-right panel is same as the upper-left panel but for $\lambda$ vs. $m_{\rm DM}$. The relatively large coupling $\lambda$ is favorable for large $R$. The lower-left panel shows that there is no specific correlation between $\lambda$ and $M_3$. Similarly, there is no specific dependence of $\lambda$ on $m_0$ as shown in the lower-right panel. From these two lower panels, there is no significant correlation between the desired large $R$ and mass of heavy fermions, $M_3$ and $m_0$. 

Figure \ref{Fig:fig_HighOmegaScoto} shows that the model parameters leading to a large $R$ cannot be strongly constrained. The large R is obtained based on the non-linear relations of the scotogenic model parameters. The only thing we can say is that the small mass of the DM and the large coupling constant are favourable for a large $R$; however, if a new fermion at $\mathcal{O}$(1) GeV and a new scalar that couples to standard Higgs bosons with $\lambda \simeq 10^{-8} - 10^{-7}$ are found in future experiments, we can expect the case of $R \ge 0.65$. Although, it cannot be said that the scotogenic model is completely correct among the numerous models of DM, we would like to suppose the scotogenic model is the correct DM model as well as the correct neutrino mass model in this study. 

\color{black}
We would like to comment on the warm DM constraints \cite{Fujita2014PRD}. According to the recent study on warm DM constraints \cite{Baldes2020JCAP}, if all the relic abundance of GeV scale DM comes from PBH evaporation, PBHs with mass $M_{\rm PBH}/M_{\rm Pl} \lesssim 10^{10}$ are not allowed by the conservative Lyman-$\alpha$ bound for warm DM. In the Ref.\cite{Kitabayashi2022PTEP1}, we observe that almost all of the relic abundance can caused by PBH evaporation in some specific parameter sets in the scotogenic model. Thus, PBH with a mass of  $M_{\rm PBH}/M_{\rm Pl} \lesssim 10^{10}$ ($M_{\rm PBH} \lesssim 2.18\times 10^5$ g)  for GeV scale DM are not allowed by the warm DM constraint in this study.

\subsection{Constraints with GWs}
\begin{figure}[t]
\begin{center}
\includegraphics[scale=1.0,pagebox=cropbox,clip]{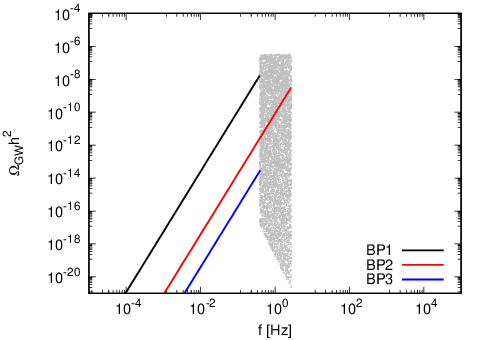}
\caption{Amplitude $\Omega_{\rm GW}$ and frequency $f$ of the GWs for the benchmark point displayed in Fig. \ref{Fig:mdm_mpbh_beta}. Peak amplitude and peak frequency of the GWs for all allowed $(m_{\rm DM}, M_{\rm PBH})$ from particle physics experiments and CMB observations displayed in Fig. \ref{Fig:mdm_mpbh_beta} are plotted (gray dots).}
\label{Fig:f_OmegaGWh2_BP} 
\end{center}
\end{figure}

\begin{figure}[t]
\begin{center}
\includegraphics[scale=1.0]{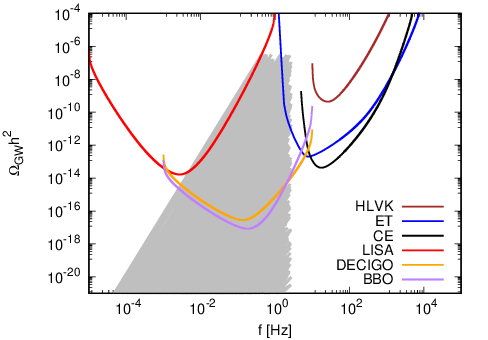}
\caption{\color{black}GWs amplitudes and frequencies consistent with the particle physics experiments and CMB observations (gray lines); lines overlapping and appear as filled-in areas. Sensitivity curves of future GW detectors are displayed for comparison. }
\label{Fig:f_OmegaGWh2} 
\end{center}
\end{figure}

\begin{figure}[t]
\begin{center}
\includegraphics[scale=1.0,pagebox=cropbox,clip]{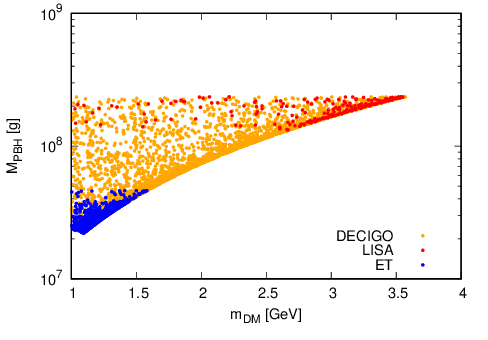}
\caption{Allowed scotogenic DM and PBH mass ranges upon detecting GWs at DECIGO (orange), LISA (red), and ET (blue).}
\label{Fig:mdm_MPBH_detector} 
\end{center}
\end{figure}

The constraints on the mass of scotogenic DM and PBH are estimated considering GWs into account.

Figure \ref{Fig:f_OmegaGWh2_BP} shows the amplitude $\Omega_{\rm GW}$ and frequency $f$ of the GWs with the peak frequency and peak amplitude 
\begin{equation}
(f^{\rm peak}\rm{[Hz]},\Omega_{\rm GW}^{\rm peak})
=
\begin{cases}
(0.387, 3.70\times 10^{-8}) &  {\rm BP1} \\
(2.68, 6.81\times 10^{-9}) & {\rm BP2} \\
(0.400, 6.41\times 10^{-14}) &  {\rm BP3} \\
\end{cases}
\end{equation}
at the benchmark point 
\begin{equation}
(m_{\rm DM} [{\rm GeV}], M_{\rm PBH}  [10^{7} {\rm g}])
=
\begin{cases}
(3.56, 23.5) & \ {\rm BP1} \\
(1.06, 2.26) & \ {\rm BP2} \\
(1.04, 23.4) & \ {\rm BP3} \\
\end{cases}
\end{equation}
shown in Fig. \ref{Fig:mdm_mpbh_beta}. The peak amplitude and peak frequency of the GWs are plotted for all allowed $(m_{\rm DM}, M_{\rm PBH})$ from particle physics experiments and CMB observations (Fig. \ref{Fig:mdm_mpbh_beta}; gray dots).  

Recall that the maximum frequency of GWs $f^{\rm peak}$ is determined by $M_{\rm PBH}$ from Eq. (\ref{Eq:fpeak}). Corresponding to the narrow $M_{\rm PBH}$ range in Fig. \ref{Fig:mdm_mpbh_beta}, the allowed maximum frequency $f^{\rm peak}$ shown in Fig. \ref{Fig:f_OmegaGWh2_BP} is limited to a narrow range as
\begin{eqnarray}
0.387 \lesssim  & \ f^{\rm peak}\rm{[Hz]} \ & \lesssim 2.76, \nonumber \\
2.21\times 10^{-21} \lesssim & \Omega_{\rm GW}^{\rm peak} & \lesssim 3.25\times 10^{-7}.
\end{eqnarray}

Figure \ref{Fig:f_OmegaGWh2} illustrates the range of GW amplitudes and frequencies consistent with the particle experiments and CMB observations, represented by gray lines (which overlap to appear as filled-in areas). Not all data from numerical calculations are displayed in the figure; they are selectively thinned to minimize the electronic file size. Additionally, the sensitivity curves of future GW detectors are presented for comparative analysis. These sensitivity curves are plotted based on data from Ref. \cite{Schmitz2021JHEP}.

Figure \ref{Fig:f_OmegaGWh2} indicate that the GWs observable by LISA and ET are restricted to a narrow frequency range. We will elucidate that the future detection of GWs by these instruments could impose stringent constraints on the masses of scotogenic DM and PBHs, constraints that are hitherto unattained.

Note that in Fig. \ref{Fig:f_OmegaGWh2}, the GWs under consideration in this study are precluded from existing within the sensitivity regions of the HLVK and the CE. Nevertheless, a plethora of GW sources, other than PBH density fluctuations in the PBH-dominated universe, do exist. Thus, even if GWs were detected in future observations by HLVK and CE, the proposed scenario in this study would not be immediately invalidated. Additionally, GW components originating from sources other than PBH density fluctuations might coexist in the GW region studied herein, rendering the obtained constraints to be conservative.

Figure \ref{Fig:mdm_MPBH_detector} illustrates the allowed mass ranges for scotogenic DM and PBHs when GWs are detected by DECIGO (depicted in orange), LISA (in red), and ET (in blue). The allowed mass range for BBO  closely resembles that for DECIGO and is, therefore, omitted for brevity. In case the GWs are detected by DECIGO, the allowed masses for scotogenic DM and PBHs can be expressed as
\begin{eqnarray}
\begin{aligned}
1.00 \lesssim  m_{\rm DM} [{\rm GeV}] \lesssim 3.57, \\
2.22  \lesssim  M_{\rm PBH}  [10^{7} {\rm g}] \lesssim 23.5,
\end{aligned}
\qquad {\rm (DECIGO)},
\end{eqnarray}
which refers to the same constraint derived from particle physics experiments and CMB observations in Eq.(\ref{Eq:constraint_wo_GWs}). Therefore, GWs observations by DECIGO will not provide stringent constraints on the masses of scotogenic DM and PBH.

This DECIGO situation alters dramatically for LISA and ET. The allowed scotogenic DM and PBH mass range can be revised as
\begin{eqnarray}
\begin{aligned}
1.03 \lesssim  m_{\rm DM} [{\rm GeV}] \lesssim 3.55, \\
13.2  \lesssim  M_{\rm PBH}  [10^{7} {\rm g}] \lesssim 23.5, 
\end{aligned}
\qquad {\rm (LISA)}
\end{eqnarray}
when GWs are detected in LISA. Only the relatively large PBH masses, around $10^8$ g, are allowed. By contrast, when GWs are detected by ET, the allowed masses are 
\begin{eqnarray}
\begin{aligned}
1.00 \lesssim  m_{\rm DM} [{\rm GeV}] \lesssim 1.58, \\
2.22  \lesssim  M_{\rm PBH}  [10^{7} {\rm g}] \lesssim 4.61, 
\end{aligned}
\qquad {\rm (ET)}.
\end{eqnarray}
Only the extremely limited regions in which both DM and PBH masses are relatively small are allowed.

\section{Summary\label{section:summary}}

The scotogenic DM is a widely studied DM candidate that is produced by thermal processes as well as by the evaporation of PBHs. The mass of scotogenic DM and PBHs could be constrained by particle physics experiments and CMB observations. 

Recent reports indicated that the curvature fluctuations of PBHs at the PBH-dominated epoch in the early universe can be a source of the so-called induced GWs. 

In this study, we demonstrated that stringent constraints on the mass of scotogenic DM and PBHs can be obtained from the detection of the induced GWs in LISA and ET. Especially, when GWs are detected by ET, the allowed region of mass of scotogenic DM and PBH are stringently limited at $m_{\rm DM} \simeq \mathcal{O}(1)$ GeV and $M_{\rm PBH} \simeq \mathcal{O} (10^7)$ g.  







\end{document}